\documentclass[a4paper, 10pt, conference]{ieeeconf}%
\usepackage{amsmath}
\usepackage{amsfonts}
\usepackage{amssymb}
\usepackage{graphicx}%
\IEEEoverridecommandlockouts
\newtheorem{theorem}{Theorem}

\newtheorem{corollary}[theorem]{Corollary}

\newtheorem{remark}[theorem]{Remark}

\begin{document}

\title{$K$-User Fading Interference Channels: The Ergodic Very Strong Case}
\author{Lalitha Sankar, Jan Vondrak, and H. Vincent Poor \thanks{This research was
supported in part by the Air Force Office of Scientific\ Research under Grant
FA9550-08-1-0480 and in part by fellowships from the Princeton University
Council on\ Science and Technology. }\thanks{L. Sankar and H. V. Poor are with
the Department of Electrical Engineering, Princeton University, Princeton, NJ,
USA. J. Vondrak is with IBM, Almaden, CA, USA. The work was done when he was
with the Department of Mathematics, Princeton University.
\{lalitha,poor@princeton.edu,jvondrak@gmail.com\texttt{{\small \}}}}}
\maketitle

\begin{abstract}
Sufficient conditions required to achieve the interference-free capacity
region of ergodic fading $K$-user interference channels (IFCs) are obtained.
In particular, this capacity region is shown to be achieved when every
receiver decodes all $K$ transmitted messages such that the channel statistics
and the waterfilling power policies for all $K$ (interference-free) links
satisfy a set of $K(K-1)$ \textit{ergodic very strong }conditions. The result
is also of independent interest in combinatorics.

\end{abstract}

\thispagestyle{empty} \pagestyle{empty}



\section{Introduction}

The $K$-user interference channel (IFC) is a network with $K$
transmitter-receiver pairs (also referred to as users or links) in which each
transmitter transmits to its intended receiver while creating interference at
one or more of the unintended receivers. In general, the problem of
determining the capacity region of a $K$-user IFC remains open. Capacity
regions are known for the two-user \textit{strong} IFC
\cite{cap_theorems:Sato_IC,cap_theorems:KobayashiHan_IC} and the \textit{very
strong} IFC \cite{cap_theorems:Carleial_VSIFC}, and in both cases the capacity
region is achieved when both receivers decode both the intended and
interfering signals, i.e., the IFC reduces to a compound MAC (C-MAC)
\cite{cap_theorems:Ahlswede_CMAC}. The \textit{very strong} IFC is a sub-class
of the class of strong IFCs for which the sum-capacity and the capacity region
are determined by the interference-free bottleneck links from the two
transmitters to their intended receivers. On the other hand, only the
sum-capacity is known for a class of \textit{weak }one-sided non-fading
two-user IFCs \cite{cap_theorems:Costa_IC} and is achieved by ignoring
interference (i.e., treating it as noise). More recently, for the two-sided
model, the sum-capacity of a class of noisy or very weak Gaussian IFCs is
determined independently in \cite{cap_theorems:ShangKramerChen},
\cite{cap_theorems:MotaKhan}, and \cite{cap_theorems:AR_VVV} and these results
have been extended for $K>2$ in \cite{cap_theorems:AR_VVV} and
\cite{cap_theorems:Shang_ISIT}. \ 

\bigskip

Ergodic fading and parallel\ Gaussian IFCs model the fading properties of
wireless networks. In this paper, we focus on $K$-user ergodic fading Gaussian
IFCs and seek to determine a set of conditions for which the sum-capacity of
$K$ interference-free links can be achieved. Recently, in
\cite{cap_theorems:LSXSEEVP} which develops sum-capacity and separability
results for two-user ergodic fading Gaussian IFCs, an \textit{ergodic very
strong }(EVS) sub-class has been identified as a collection of ergodic fading
Gaussian IFCs with a weighted mixture of weak and strong sub-channels (fading
states) for which the sum of the interference-free capacities of the two user
links can be achieved. Two-user parallel Gaussian IFCs have also been studied
in \cite{cap_theorems:ChuCioffi_IC,cap_theorems:Shang_03} and
\cite{cap_theorems:KaistParIFC}. For fading IFCs with three or more users,
\cite{cap_theorems:CadamJafar_IFCAlign} presents an \textit{interference
alignment} scheme to show that the sum-capacity of a $K$-user IFC scales
linearly with $K$ in the high signal-to-noise ratio (SNR) regime when all
links in the network have similar channel statistics.

\bigskip

The sum-capacity and capacity region of two-user EVS IFCs are achieved when
each user transmits to its intended receiver as if there were two independent
interference-free links. The sum-capacity optimal power policies are thus the
classic point-to-point waterfilling solutions developed in
\cite{cap_theorems:GoldsmithVaraiya}. While the sum-rate achieved thus is
always an outer bound on the sum-capacity of IFCs, in
\cite{cap_theorems:LSXSEEVP} it is shown that this outer bound can be achieved
when both receivers decode both messages, i.e., the IFC converts to a C-MAC,
provided the sum of the interference-free capacities of each link is strictly
smaller than the multiaccess sum-rates achieved at each receiver. These
sufficient conditions does not impose strong or weak conditions on any
sub-channel and only involve fading averaged conditions on the waterfilling policies.

\bigskip

For two-user IFCs, the above-mentioned sufficient conditions are obtained
simply by enumerating all possible intersections of the two MAC pentagons and
identifying the intersection satisfying the EVS definition. They can also be
obtained using the fact that the multiaccess rate region at each receiver is a
polymatroid \cite{cap_theorems:TH01}, and therefore, a single known lemma on
the sum-rate of two intersecting polymatroids \cite[chap. 46]%
{cap_theorems:Schrijver01} readily yields a closed-form expression for each
possible intersection of the two MAC regions. However, for intersections of
three or more polymatroids no such lemma exists that can simplify the
sum-capacity analysis of C-MACs with $K$ transmitters and $K$ receivers. This
in turn makes it difficult to identify a set of sufficient conditions for
EVS\ IFCs when every receiver is allowed to decode the message from every transmitter.

\bigskip

In this paper, we determine a set of sufficient conditions for which the
sum-capacity is the sum of the capacities of $K$ interference-free links when
all receivers are allowed to decode the messages from all transmitters. This
in turn corresponds to determining the conditions for which the intersection
of $K$ rate polymatroids results in a $K$-dimensional \textit{box}
(hyper-cube) that is uniquely defined by the interference-free point-to-point
rates of the $K$ links. We show that the number of sufficient conditions for
EVS\ IFCs grows quadratically in $K$. As a special case, for two-user IFCs, we
show that these conditions are both necessary and sufficient.

\bigskip

Recently, in \cite{cap_theorems:Sridharan_01}, the authors present sufficient
conditions for which a $K$-user symmetric non-fading IFC achieves the
sum-capacity of $K$ interference-free links. The conditions are developed
using lattice codes which enables complete decoding of the interference but
not the message from every interfering transmitter. Using the fact that the
non-fading IFC is a special case of an EVS\ IFC, we compare our results to
those in \cite{cap_theorems:Sridharan_01} and show that decoding messages from
all users is best used in the low power regime or when the symmetric
cross-links are relatively closer to unity while lattice codes are
advantageous otherwise. Finally, we note that the results developed here are
also of independent interest in combinatorics.

\bigskip

The paper is organized as follows. We present channel model and preliminaries
in Section \ref{Sec_II}. The main result and the proof are developed in
Section \ref{Sec_III}. We discuss the results and present numerical examples
in\ Section \ref{Sec_IV}. We conclude in Section \ref{Sec_V}.%

\begin{figure}
[ptb]
\begin{center}
\includegraphics[
height=1.8645in,
width=2.8599in
]%
{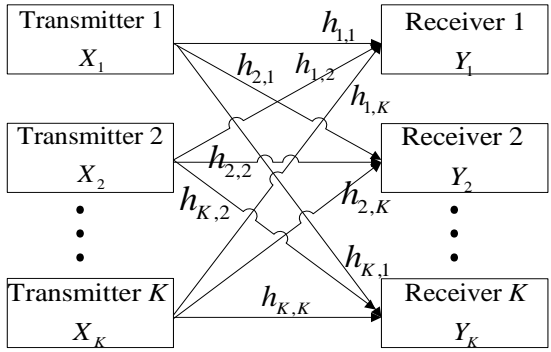}%
\caption{A $K$-user ergodic fading interference channel.}%
\label{Fig_IFCKUser}%
\end{center}
\end{figure}

\section{\label{Sec_II}Channel Model}

A $K$-user (or $K$-link) ergodic fading Gaussian IFC consists of $K$
transmitter-receiver pairs, each link pair indexed by $k$, $k\in
\mathcal{K}=\left\{  1,2,\ldots,K\right\}  $, as shown in Fig.
\ref{Fig_IFCKUser}. Transmitter $k$ uses the channel $n$ times to transmit its
message $W_{k}$, which is distributed uniformly in the set $\,\{1,2,\ldots
,2^{B_{k}}\}$ and is independent of the messages from the other transmitters,
to its intended receiver $k$, at a rate $R_{k}=B_{k}/n$ bits per channel use.
In each use of the channel, transmitter $k$ transmits the signal $X_{k}$ while
receiver $k$ receives $Y_{k}$, $k\in\mathcal{K}.$ For $\mathbf{X}=\left[
X_{1}\text{ }X_{2}\text{ }\ldots\text{ }X_{K}\right]  ^{T}$, the channel
output vector $\mathbf{Y}=\left[  Y_{1}\text{ }Y_{2}\text{ }\ldots\text{
}Y_{K}\right]  ^{T}$ in a single channel use is given by%
\begin{equation}
\mathbf{Y}=\mathbf{HX}+\mathbf{Z} \label{IC_Y}%
\end{equation}
where $\mathbf{Z}=\left[  Z_{1}\text{ }Z_{2}\text{ }\ldots\text{ }%
Z_{K}\right]  ^{T}$ is a noise vector with entries that are zero-mean, unit
variance, circularly symmetric complex Gaussian noise variables and
$\mathbf{H}$ is a random matrix of fading gains with entries $H_{m,k}$, for
all $m,k\in\mathcal{K}$, such that $H_{m,k}$ denotes the fading gain between
receiver $m$ and transmitter $k$. We use $\mathbf{h}$ to denote a realization
of $\mathbf{H}$. We assume the fading process $\left\{  \mathbf{H}\right\}  $
is stationary and ergodic but not necessarily Gaussian. Note that the channel
gains $H_{m,k}$, for all $m$ and $k$, are not assumed to be independent;
however, $\mathbf{H}$ is assumed to be known instantaneously at all the
transmitters and receivers.

\bigskip

Over $n$ uses of the channel, the transmit sequences $\left\{  X_{k,i}%
\right\}  $ are constrained in power according to%
\begin{equation}
\left.  \sum\limits_{i=1}^{n}\left\vert X_{k,i}\right\vert ^{2}\leq
n\overline{P}_{k}\right.  ,\text{ for }k\in\mathcal{K}\text{.} \label{IFC_Pwr}%
\end{equation}
Since the transmitters know the fading states of the links on which they
transmit, they can allocate their transmitted signal power according to the
channel state information. A power policy \underline{$P$}$(\mathbf{h})$ with
entries $P_{k}(\mathbf{h})$ for $k\in\mathcal{K}$ is a mapping from the fading
state space consisting of the set of all fading states (instantiations)
$\mathbf{h}$ to the set of non-negative real values in $\mathcal{R}_{+}^{K}$.
We write \underline{$P$}$(\mathbf{H})$ to describe explicitly the policy for
the entire set of random fading states. For an ergodic fading channel,
(\ref{IFC_Pwr}) then simplifies to
\begin{equation}
\left.  \mathbb{E}\left[  P_{k}(\mathbf{H})\right]  \leq\overline{P}%
_{k}\right.  \text{ for }k\in\mathcal{K}, \label{ErgPwr}%
\end{equation}
where the expectation in (\ref{ErgPwr}) is taken over the distribution of
$\mathbf{H}$.

\bigskip

For the special case in which all receivers decode the messages from all
transmitters, we obtain a compound MAC. We write $\mathcal{C}_{\text{IFC}%
}\left(  \underline{\overline{P}}\right)  $ and $\mathcal{C}_{\text{C-MAC}%
}\left(  \underline{\overline{P}}\right)  $ to denote the capacity regions of
an ergodic fading IFC and C-MAC, respectively, where $\underline{\overline{P}%
}$ is a vector whose entries are the average power constraints $\overline
{P}_{k}$, for $k\in\mathcal{K}$. Our definitions of average error
probabilities, capacity regions, and achievable rate pairs $\left(
R_{1},R_{2},\ldots,R_{K}\right)  $ for both the IFC and C-MAC mirror the
standard information-theoretic definitions \cite[Chap. 14]%
{cap_theorems:CTbook}.

\bigskip

Throughout the sequel, we use the terms fading states and sub-channels
interchangeably. $C(x)$ denotes $\log(1+x)$ where the logarithm is to the base
2 and $R_{\mathcal{S}}$ denotes $%
{\textstyle\sum\nolimits_{k\in\mathcal{S}}}
R_{k}$ for any ${\mathcal{S}}$ $\subseteq\mathcal{K}$. We assume that the
reader is familiar with sub-modular functions and polymatroids (see, for
example, \cite{cap_theorems:Schrijver01}).

\section{\label{Sec_III}Achieving the Interference-free Capacity Region}

The following theorem summarizes the main result of this paper.

\begin{theorem}
\label{Th_EVS}A $K$-user ergodic fading IFC achieves the interference-free
capacity region of $K$ independent links if the waterfilling solutions
$P_{k}^{\left(  wf\right)  }\left(  H_{kk}\right)  $ for the
(interference-free) ergodic fading point-to-point links between transmitters
$k$ and receivers $k$, $k=1,2,\ldots,K$, satisfy%
\begin{multline}
\mathbb{E}\left[  C\left(  \left\vert H_{k,k}\right\vert ^{2}P_{k}^{\left(
wf\right)  }\left(  H_{kk}\right)  \right)  \right]  <C_{sum}^{\left(
j\right)  }\left(  \mathcal{K}\right) \label{EVS_Cond}\\
-C_{sum}^{\left(  j\right)  }\left(  \mathcal{K}\backslash\left\{  k\right\}
\right)  ,\text{for all \thinspace}j,k\in\mathcal{K},j\not =k,
\end{multline}
where for any $\mathcal{A}\subseteq\mathcal{K}$%
\begin{equation}
C_{sum}^{\left(  j\right)  }\left(  \mathcal{A}\right)  =\mathbb{E}\left[
C\left(  \sum\limits_{m\in\mathcal{A}}\left\vert H_{j,m}\right\vert ^{2}%
P_{m}^{\left(  wf\right)  }\left(  H_{kk}\right)  \right)  \right]  \text{.}%
\end{equation}
The capacity region of the resulting ergodic very strong IFC is
\begin{align}
\mathcal{C}_{\text{IFC}}^{EVS}  &  =\left\{  \left(  R_{1},R_{2},\ldots
,R_{K}\right)  :\right. \nonumber\\
&  \text{ }\left.  R_{k}\leq\mathbb{E}\left[  C\left(  \left\vert
H_{k,k}\right\vert ^{2}P_{k}^{wf}\left(  H_{k,k}\right)  \right)  \right]
,k\in\mathcal{K}\right\}  \label{EVS_CapR}%
\end{align}
and the sum-capacity is
\begin{equation}
\sum_{k=1}^{K}\mathbb{E}\left[  C\left(  \left\vert H_{k,k}\right\vert
^{2}P_{k}^{wf}\left(  H_{k,k}\right)  \right)  \right]  . \label{EVS_SC}%
\end{equation}

\end{theorem}

\begin{remark}
The conditions in (\ref{EVS_Cond}) involve averaging over all channel states
and do not require every sub-channel to be strong. As with the two-user
ergodic fading IFCs, the capacity achieving scheme for $K$-user EVS\ IFCs
requires coding jointly across all sub-channels. For $K=2$, (\ref{EVS_Cond})
simplifies to the EVS conditions in \cite[Theorem 2]{cap_theorems:LSXSEEVP}.
The conditions for a $K$-user non-fading very strong IFC are simply a special
case of (\ref{EVS_Cond}) obtained for a constant $\mathbf{H}$.
\end{remark}

\begin{remark}
The conditions in (\ref{EVS_Cond}) are equivalent to the requirements that the
rate achieved by each transmitter in the presence of interference from all
other users at each of the unintended receivers is at least as large as the
interference-free rate achieved at its intended receiver.
\end{remark}

\begin{corollary}
\label{Cor1}For a class of symmetric non-fading IFCs with $H_{k,k}=1$,
$H_{j,k}=a$ for all $j,k\in\mathcal{K},$ $j\not =k$, and $\overline{P}%
_{k}=\overline{P}$, (\ref{EVS_Cond}) reduces to the condition%
\begin{equation}
\overline{P}<\frac{a^{2}-1}{1+\left(  K-2\right)  a^{2}} \label{Pmax_symIFC}%
\end{equation}
or equivalently%
\begin{equation}
a^{2}>\frac{1+\overline{P}}{1-\left(  K-2\right)  \overline{P}}.
\label{VSSymm_a}%
\end{equation}
Thus, for any $a>0$ and $K>2$, $\overline{P}<1$; furthermore, for large $K$,
$\overline{P}$ scales inversely with $K$. Conversely, for large $K$, $a^{2}$
scales linearly with $K$.
\end{corollary}

\begin{remark}
A very strong condition for $K$-user symmetric IFCs is presented in \cite[eqn.
(5)]{cap_theorems:Sridharan_01} which requires that $a^{2}$ grow exponentially
with $K$ when each receiver decodes all the unintended messages before
decoding its intended message. It is unclear whether the condition in
\cite[eqn. (5)]{cap_theorems:Sridharan_01} ensures that the intersection of
the MAC polymatroids, one at each receiver, is a box. In contrast, the
condition in (\ref{VSSymm_a}) only grows linearly in $K$ and ensures a box intersection.
\end{remark}

\begin{remark}
\label{Rem_Lat}\cite{cap_theorems:Sridharan_01} also presents a sufficient
condition using lattice codes for interference-free communications in
symmetric $K$-user IFCs as $a^{2}>\left(  \overline{P}+1\right)
^{2}/\overline{P}$ which is independent of the number of users.
\end{remark}

\begin{proof}
\textit{Outer Bound}: An outer bound on the sum-capacity of an IFC results
from eliminating interference at all the receivers thereby reducing it to $K$
interference-free point-to-point links. From \cite[Appendix]%
{cap_theorems:GoldsmithVaraiya}, the capacity achieving policy for each link
requires each transmitter to waterfill over its fading link to its receiver,
and thus, we have that any achievable rate tuple $\left(  R_{1},R_{2}%
,\ldots,R_{K}\right)  $ must satisfy
\begin{equation}
\sum_{k=1}^{K}R_{k}\leq\sum_{k=1}^{K}\mathbb{E}\left[  C\left(  \left\vert
H_{k,k}\right\vert ^{2}P_{k}^{wf}\left(  H_{k,k}\right)  \right)  \right]  .
\label{VS_OB}%
\end{equation}

\textit{Inner Bound}: Consider the achievable scheme in which every receiver
decodes all the interfering signals, i.e., the IFC is converted to a C-MAC.
Assuming every transmitter encodes its message across all sub-channels and
every receiver jointly decodes all messages across all sub-channels, the
Gaussian MAC rate region achieved at receiver $k$ when the power policy at
transmitter $m$ is $P_{m}^{\left(  wf\right)  }$, for all $k,m\in\mathcal{K}$
is given by \cite[Theorem 1]{cap_theorems:LSXSEEVP}
\begin{multline}
\mathcal{R}_{k}\left(  \underline{P}^{\left(  wf\right)  }\right)  =\left\{
\left(  R_{1},R_{2},\ldots,R_{K}\right)  :R_{\mathcal{S}}\leq f_{k}^{\ast
}\left(  \mathcal{S}\right)  ,\right. \label{MAC_k}\\
\text{ }\left.  \text{for all }\mathcal{S}\subseteq\mathcal{K}\right\}
\end{multline}
where
\begin{equation}
f_{k}^{\ast}\left(  \mathcal{S}\right)  =\mathbb{E}\left[  C\left(
{\textstyle\sum\nolimits_{m\in\mathcal{S}}}
\left\vert H_{k,m}\right\vert ^{2}P_{m}^{wf}\left(  H_{m,m}\right)  \right)
\right]  .
\end{equation}
It can be easily verified that the functions $f_{k}^{\ast}\left(
\mathcal{S}\right)  $, for all $k$, are sub-modular functions and the MAC rate
regions $\mathcal{R}_{k}\left(  \underline{P}^{\left(  wf\right)  }\right)  $
are polymatroids (see for e.g., \cite{cap_theorems:LSYLNMHVP}). For any
$\mathcal{S},\mathcal{A}\subseteq\mathcal{K}$ such that $\mathcal{S}%
\cap\mathcal{A=\emptyset}$, we write $f_{k}^{\ast}\left(  \mathcal{S}%
|\mathcal{A}\right)  $ as
\begin{equation}
f_{k}^{\ast}\left(  \mathcal{S}|\mathcal{A}\right)  \equiv\mathbb{E}\left[
C\left(  \frac{%
{\textstyle\sum\nolimits_{m\in\mathcal{S}}}
\left\vert H_{k,m}\right\vert ^{2}P_{m}^{wf}\left(  H_{m,m}\right)  }{1+%
{\textstyle\sum\nolimits_{m\in\mathcal{A}}}
\left\vert H_{k,m}\right\vert ^{2}P_{m}^{wf}\left(  H_{m,m}\right)  }\right)
\right]  , \label{fSgivA}%
\end{equation}
i.e., $f_{k}^{\ast}\left(  \mathcal{S}|\mathcal{A}\right)  $ is the rate
achieved by the users in $\mathcal{S}$ at receiver $k$ in the presence of
interference from the users in a disjoint set $\mathcal{A}$.

We now show that when (\ref{EVS_Cond}) is satisfied the intersection of
$\mathcal{R}_{k}\left(  \underline{P}^{\left(  wf\right)  }\right)  $ for all
$k$ is a $K$-dimensional hyper-cube. To this end, for ease of analysis, we
first write (\ref{EVS_Cond}) in terms of $f_{k}^{\ast}(\mathcal{K)}$ as
\begin{equation}
f_{k}^{\ast}\left(  \left\{  k\right\}  \right)  <f_{j}^{\ast}\left(
\mathcal{K}\right)  -f_{j}^{\ast}\left(  \mathcal{K}\backslash\left\{
k\right\}  \right)  ,\text{ for all }j,k\in\mathcal{K},j\not =k.\label{EVSCd}%
\end{equation}
Thus, given (\ref{EVSCd}), we now prove that
\begin{equation}%
\begin{array}
[c]{cc}%
\sum\limits_{k\in\mathcal{S}}f_{k}^{\ast}\left(  \left\{  k\right\}  \right)
\leq f_{j}^{\ast}\left(  \mathcal{S}\right)  , & \text{for all }j\text{ and
}\mathcal{S}\subseteq\mathcal{K}.
\end{array}
\label{EVS_eq}%
\end{equation}
Without loss of generality, let $j=1$ and $\mathcal{S}=\mathcal{K}$. Thus, we
have
\begin{subequations}
\label{fBound}%
\begin{align}
\sum\limits_{k\in\mathcal{K}}f_{k}^{\ast}\left(  \left\{  k\right\}  \right)
&  =f_{1}^{\ast}\left(  \left\{  1\right\}  \right)  +f_{2}^{\ast}\left(
\left\{  2\right\}  \right)  +\ldots+f_{K}^{\ast}\left(  \left\{  K\right\}
\right)  \\
&  \leq f_{1}^{\ast}\left(  \left\{  1\right\}  \right)  +f_{1}^{\ast}\left(
\left\{  1,2\right\}  \right)  -f_{1}^{\ast}\left(  \left\{  1\right\}
\right)  \label{fineq}\\
&  +f_{1}^{\ast}\left(  \left\{  1,2,3\right\}  \right)  -f_{1}^{\ast}\left(
\left\{  1,2\right\}  \right)  +\ldots\nonumber\\
&  +f_{1}^{\ast}\left(  \mathcal{K}\right)  -f_{1}^{\ast}\left(
\mathcal{K}\backslash\left\{  k\right\}  \right)  \\
&  =f_{1}^{\ast}\left(  \mathcal{K}\right)
\end{align}
where (\ref{fineq}) follows from (\ref{EVSCd}) and the fact that for any
$\mathcal{S\subset K}$ such that $k\not \in \mathcal{S}$, using chain rule for
mutual information, we have%
\end{subequations}
\begin{align}
&  f_{j}^{\ast}\left(  \mathcal{K}\right)  -f_{j}^{\ast}\left(  \mathcal{K}%
\backslash\left\{  k\right\}  \right)  \nonumber\\
&  =f_{j}^{\ast}\left(  \mathcal{S\cup}\left\{  k\right\}  \right)
+f_{j}^{\ast}\left(  \mathcal{K}\backslash\left(  \mathcal{S}\cup\left\{
k\right\}  \right)  |\mathcal{S}\cup\left\{  k\right\}  \right)  \nonumber\\
&  -f_{j}^{\ast}\left(  \mathcal{S}\right)  -f_{j}^{\ast}\left(
\mathcal{K}\backslash\left(  \mathcal{S}\cup\left\{  k\right\}  \right)
|\mathcal{S}\right)  \label{f_ineq1}\\
&  \leq f_{j}^{\ast}\left(  \mathcal{S\cup}\left\{  k\right\}  \right)
-f_{j}^{\ast}\left(  \mathcal{S}\right)  \label{f_ineq}%
\end{align}
where (\ref{f_ineq}) follows from the fact that due to additional interference
from user $k,$ the second term to the right of the equality in (\ref{f_ineq1})
is smaller than the fourth term where all terms in (\ref{f_ineq1}) can be
expanded using (\ref{fSgivA}).

Following steps similar to (\ref{fBound}), one can show that $%
{\textstyle\sum\nolimits_{k\in\mathcal{S}}}
f_{k}^{\ast}\left(  \left\{  k\right\}  \right)  \leq f_{1}^{\ast}\left(
\mathcal{S}\right)  $ for all $\mathcal{S}\subset\mathcal{K}.$ Furthermore,
the same steps can also be used to show that (\ref{EVS_eq}) holds for all
$j\in\mathcal{K}.$ Let $R_{k}^{\ast}=f_{k}^{\ast}\left(  \left\{  k\right\}
\right)  $ for all $k$. Thus, from (\ref{EVS_eq}) and (\ref{MAC_k}), we have
that
\begin{align}
&
\begin{array}
[c]{cc}%
\left(  R_{1}^{\ast},R_{2}^{\ast},\ldots,R_{K}^{\ast}\right)  \in
\mathcal{R}_{j}\left(  \underline{P}^{\left(  wf\right)  }\right)  , &
\text{for all }j\in\mathcal{K},
\end{array}
\\
&  \Longrightarrow\left(  R_{1}^{\ast},R_{2}^{\ast},\ldots,R_{K}^{\ast
}\right)  \in\cap_{j=1}^{K}\mathcal{R}_{j}\left(  \underline{P}^{\left(
wf\right)  }\right)  .
\end{align}
Since the intersection of $K$ orthogonal rate planes $R_{k}^{\ast}$ yields a
box (a hyper-cube), the C-MAC sum-capacity when (\ref{EVS_Cond}) holds is
given by (\ref{EVS_SC}). Combining this achievable sum-rate with the outer
bounds in (\ref{VS_OB}), we have that (\ref{EVS_SC}) is also the sum-capacity
of an EVS\ IFC for which the channel statistics and optimal power policy
satisfy (\ref{EVS_Cond}). Finally, since the sum-capacity also achieves the
interference-free capacity of each user, the capacity region of EVS\ IFCs is
given by (\ref{EVS_CapR}).
\end{proof}

%

\begin{figure*}[tbp] \centering
{\includegraphics[
trim=0.134085in 0.000000in 0.269816in 0.000000in,
height=3.0485in,
width=6.1281in
]%
{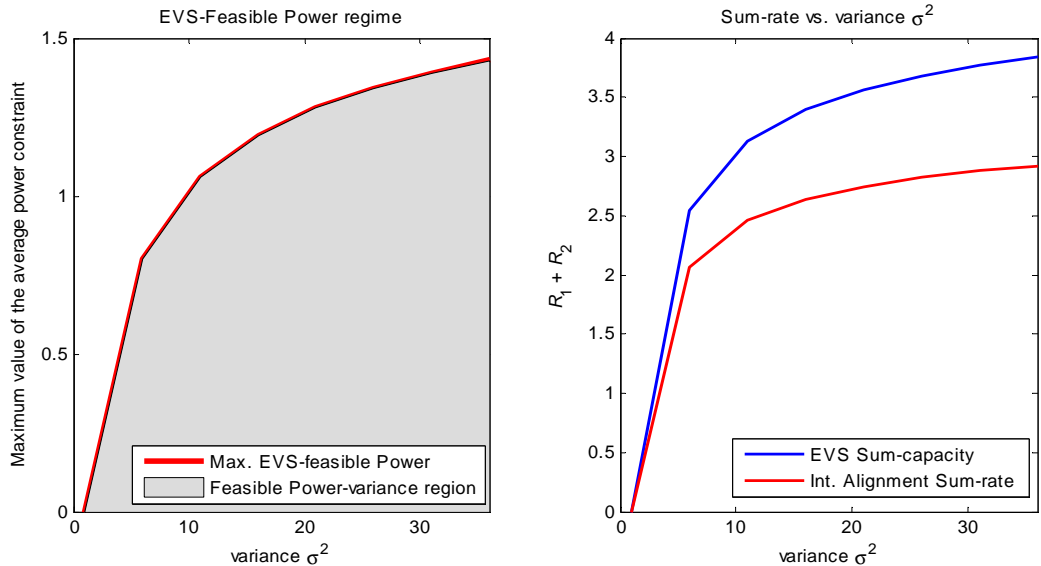}%
}%
\caption{EVS feasible power-variance region and sum-capacity.}\label{Fig_EVSplot}%
\end{figure*}%

\section{\label{Sec_IV}Discussion}

Theorem \ref{Th_EVS} summarizes a set of sufficient conditions for which the
interference-free capacity of $K$ transmit-receive pairs can be achieved when
every receiver decodes the messages from all transmitters, an achievable
scheme we henceforth refer to as the C-MAC scheme. In this section, we present
numerical examples of fading and non-fading IFCs and the feasible power and
channel gains regime for which the EVS\ IFC conditions in (\ref{EVS_Cond}) are satisified.

\bigskip

We first consider a three-user ergodic fading IFC with non-fading unit-gain
direct links and cross-links that are independent and identically distributed
Rayleigh faded links, i.e., $H_{j,k}\sim\mathcal{CN}\left(  0,\sigma
^{2}\right)  $ for all $j\not =k,j,k\in\left\{  1,2,3\right\}  ,$ and
$\overline{P}_{k}=\overline{P}$ for all $k$. The resulting channel is a mix of
weak and strong sub-channels for which each user transmits at $\overline{P}$
in every sub-channel if the EVS conditions in (\ref{EVS_Cond}) are satisfied.
The feasible $\overline{P}$ vs. $\sigma^{2}$ region and the maximum
$\overline{P}\left(  \sigma^{2}\right)  $ for which (\ref{EVS_Cond}) holds in
plotted in Fig. \ref{Fig_EVSplot}(a). In Fig. \ref{Fig_EVSplot}(b), the EVS
sum-capacity when each user transmits at the maximum $\overline{P}\left(
\sigma^{2}\right)  $ is plotted as a function of the fading variance
$\sigma^{2}$. For this $\overline{P}\left(  \sigma^{2}\right)  $, also plotted
in Fig. \ref{Fig_EVSplot}(b) is the sum-rate achieved by ergodic interference
alignment in which knowledge of the channel states is used by the transmitters
to enable the cancellation of interference at all receivers simultaneously
\cite{cap_theorems:Nazer01}. As shown in both subplots, as the variance of the
cross-links increases, thereby increasing the probability of strong fading
states, the largest $\overline{P}$, and hence the sum-capacity, for which
the\ EVS\ sum-capacity is achievable also increases.

\bigskip

\begin{figure}
[ptb]
\begin{center}
\includegraphics[
trim=0.319397in 0.058724in 0.349639in 0.170096in,
height=2.8548in,
width=3.461in
]%
{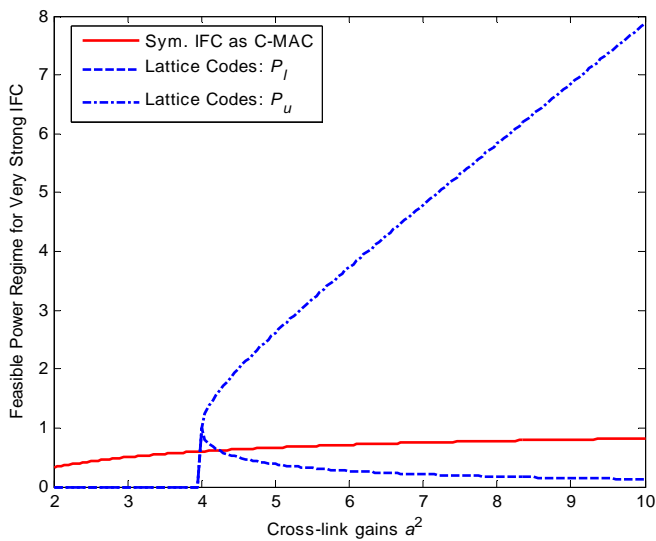}%
\caption{Bounds on feasible power vs. cross-link gains for C-MAC- and lattice
codes-based achievable schemes.}%
\label{Fig_asqrP}%
\end{center}
\end{figure}

Next we consider a non-fading three-user symmetric IFC with unit gains on the
intended links, a real positive channel gain $a$ on the cross-links, and
$\overline{P}_{k}=\overline{P}$, for $k=1,2,3$. In\ Fig. \ref{Fig_asqrP}(a),
as a function of $a^{2}$, we plot the maximum feasible $\overline{P}$ in
(\ref{Pmax_symIFC}) for which a very strong IFC results using a C-MAC
achievable scheme. As observed in\ Corollary \ref{Cor1}, we require
$\overline{P}<1$. Also included are plots of the upper $\overline{P}_{u}$ and
lower $\overline{P}_{l}$ bounds on the feasible power with lattice codes for
which a very strong IFC results. Thus, as $a^{2}$ increases, decoding the
interference using lattice codes allows a larger class of three-user symmetric
IFCs to be be considered very strong relative to decoding the message from
every user. On the other hand, only the C-MAC scheme achieves the VS condition
for $a^{2}<4$, i.e., the C-MAC achievable scheme is more appropriate in the
low-power regime in achieving the sum-capacity of $K$ interference free
point-to-point links.

A set of sufficient EVS\ conditions given by (\ref{EVS_Cond}) in Theorem
\ref{Th_EVS} prompt the question of whether these conditions are also
necessary, i.e., whether the intersection of $K$ polymatroids would cease to
be a box if one or more conditions were violated. We now present a three-user
example that shows that when all six conditions in (\ref{EVS_Cond}) for $K=3$
are not satisfied, the intersection of the $K$ MAC polymatroids is a box, i.e,
the $K$-user interference-free sum capacity can still be achieved.

\bigskip

Consider a three-user non-fading IFC with $H_{k,k}=1$ and $\overline{P}_{k}=1$
for all $k$. Thus, if the intersection of the MAC rate regions at the
receivers results in a box, each user transmits at $\overline{P}_{k}$ in every
use of the channel$.$ The cross-link gains $H_{j,k}$ for all $j,k,$
$j\not =k,$ are such that
\begin{equation}
f_{1}^{\ast}\left(  \mathcal{S}\right)  =\max_{i\in\mathcal{S}}\left(
i\right)  +0.5\left\vert \mathcal{S}\right\vert \label{fS_eg}%
\end{equation}
where $\left\vert \mathcal{S}\right\vert $ denotes the cardinality of the set
$\mathcal{S}$ and $f_{1}^{\ast}\left(  \mathcal{S}\right)  $ is obtained by
evaluating the rate bounds at $P_{k}^{\left(  wf\right)  }\left(
H_{k,k}\right)  =\overline{P}_{k}=1$ for all $k.$ Thus, from (\ref{fS_eg}),
$f_{1}\left(  \left\{  1\right\}  \right)  =1.5$, $f_{1}\left(  \left\{
2\right\}  \right)  =2.5,$ $f_{1}\left(  \left\{  3\right\}  \right)  =3.5,$
$f_{1}\left(  \left\{  1,2\right\}  \right)  =3,$ $f_{1}\left(  \left\{
1,3\right\}  \right)  =f_{1}\left(  \left\{  2,3\right\}  \right)  =4,$ and
$f_{1}\left(  \left\{  1,2,3\right\}  \right)  =4.5$.%

\bigskip

The bounds at the other two receivers are given by $f_{2}^{\ast}\left(
\mathcal{S}\right)  =$ $f_{1}^{\ast}\left(  \pi\left(  \mathcal{S}\right)
\right)  $ and $f_{3}^{\ast}\left(  \mathcal{S}\right)  =f_{1}^{\ast}\left(
\pi^{2}\left(  \mathcal{S}\right)  \right)  $ where $\pi$ is a cyclic
permutation of the indexes such that each index is decreased by $1$ in a
cyclic manner such that $\left(  1,2,3\right)  $ map to $\left(  3,1,2\right)
$ and $\pi^{2}\left(  \mathcal{S}\right)  $ is obtained by applying
$\pi\left(  \mathcal{S}\right)  $ twice. Thus, $f_{2}\left(  \left\{
2\right\}  \right)  =f_{3}\left(  \left\{  3\right\}  \right)  =f_{1}\left(
\left\{  1\right\}  \right)  =1$.

\bigskip

For the three rate regions defined thus, one can verify that none of the six
conditions in (\ref{EVSCd}) is satisfied. Furthermore, from (\ref{fS_eg}), we
have that the bounds for every rate region satisfy $f_{k}^{\ast}\left(
\mathcal{S}\right)  \geq\left\vert \mathcal{S}\right\vert $. Consider the rate
tuple $\left(  R_{1},R_{2},R_{3}\right)  =\left(  1,1,1\right)  $. This tuple
satisfies \thinspace$%
{\textstyle\sum\nolimits_{k\in\mathcal{S}}}
R_{k}=\left\vert \mathcal{S}\right\vert \leq f_{k}^{\ast}\left(
\mathcal{S}\right)  $ for all $\mathcal{S}\subseteq\mathcal{K}$ and $j=1,2,3,$
i.e., $\left(  1,1,1\right)  $ satisfies the rate constraints for each of the
three MAC rate regions and therefore lies in their intersection, i.e., the
intersection of the three rate regions is a box.

\bigskip

This is so because, for the considered example, while (\ref{EVS_Cond}) is not
satisfied for all $j$ and $k,$ $j\not =k,$ the conditions $f_{k}\left(
\left\{  k\right\}  \right)  \leq f_{j}\left(  \mathcal{S}\right)
-f_{j}\left(  \mathcal{S}\backslash\left\{  k\right\}  \right)  $ for all
$\mathcal{S}\subset\mathcal{K}$, $j\not =k,$ are satisfied and suffice to
achieve the interference-free sum-capacity. Thus, one can conclude that for
the C-MAC achievable scheme, an EVS\ IFC will not result if and only if
$f_{k}\left(  \left\{  k\right\}  \right)  $ does not satisfy all $\left(
K-1\right)  \left(  2^{K}-1\right)  $ conditions, i.e., when $f_{k}\left(
\left\{  k\right\}  \right)  >f_{j}\left(  \mathcal{S}\right)  -f_{j}\left(
\mathcal{S}\backslash\left\{  k\right\}  \right)  $ for all $j,k\in
\mathcal{K},$ $j\not =k$ and $\mathcal{S}\subseteq\mathcal{K}$. The proof
follows in a straightforward manner from showing that $\left(  f_{1}^{\ast
}\left(  1\right)  ,f_{2}^{\ast}\left(  2\right)  ,\ldots,f_{K}^{\ast}\left(
K\right)  \right)  \notin\mathcal{R}_{k}(P_{k}^{\left(  wf\right)  }\left(
H_{kk}\right)  )$, for all $k$, using steps analogous to (\ref{fBound}) in
Theorem \ref{Th_EVS}. Note that only for $K=2$ are the sufficient conditions
in (\ref{EVS_Cond}) also necessary.

\section{\label{Sec_V}Concluding Remarks}

We have obtained sufficient conditions for achieving the interference-free
sum-capacity and capacity region of a $K$-link ergodic fading IFC when all
receivers are allowed to decode the messages from all transmitters. In
particular, we have shown that an an EVS\ IFC results if the channel
statistics and the interference-free capacity optimal waterfilling policies
for all links satisfy $K\left(  K-1\right)  $ conditions. For $K=2$, we have
shown that these conditions are both necessary and sufficient. Our result that
a quadratic number of conditions suffice for the intersection of $K$
polymatroids to form a box is also of independent interest in combinatorics
where few results are known on intersection of three or more polymatroids.
Finally, our results suggest that decoding interference using schemes such as
lattice codes may impose less stringent conditions on the average power and
channel statistics.

\bibliographystyle{IEEEtran}
\bibliography{IC_refs}

\end{document}